# Intrinsic ion migration dynamics in a one-dimensional organic metal halide hybrid

Zhenqi Hua[1], Azza Ben-Akacha[2], Qingquan He[2], Tianhan Liu[1], Gillian Boyce[1], Margaret van Deventer[1], Xinsong Lin[2], Hanwei Gao[1], Biwu Ma[2, *], and Peng Xiong[1, *]

[1]Department of Physics, Florida State University, Tallahassee, Florida 32306, USA

[2] Department of Chemistry and Biochemistry, Florida State University, Tallahassee, Florida 32306, USA

*Email: bma@fsu.edu, pxiong@fsu.edu

**Abstract**

Metal halide perovskites possess many physical properties amenable to optoelectronic applications, whereas the realization of these potentials has been hampered by their environmental and electronic instabilities. The morphological and molecular low dimensional perovskites and perovskite related materials have shown much promise in enhancing the chemical stability due to their unique molecular structures. Here we report on robust and reproducible four-terminal (4T) electrical measurements in a one-dimensional (1D) organic metal halide hybrid, (R)-α-methylbenzylammonium lead triiodide ((R-α-MBA)PbI$_3$) made possible by its chemical stability. The results reveal a distinct intrinsic ion migration dynamic of single exponential, which underlies the unique 4T *I-V* characteristics. The dynamic is directly verified by real-time measurements of the transient ionic current. Our observations are consistent with photo-activation and field-assisted ion migration. The elucidated intrinsic ion dynamics may provide the physical basis for understanding and modelling the ubiquitous hysteresis in metal halides based electronic devices and new insights into the dynamics of ion migration in metal halide perovskites and hybrids in general.

Over the past decade, there has been remarkable progress in material synthesis and device optimization of optoelectronic devices based on hybrid metal halide perovskites, which has resulted in rapid improvements in their optoelectronic performances. For instance, metal halide perovskites based solar cells have achieved power conversion efficiencies (PCEs) of more than 25% in single-junction architectures and close to 30% in perovskite-silicon tandem devices,[1,2] while perovskite light-emitting diodes (LEDs), in particular the green, red, and near-infrared versions, have shown internal quantum efficiencies approaching the theoretical limit.[3–8] Despite the rapid advances in the performance and efficiency of perovskite-based photoelectronics, the chemical and electrical stability against ambient condition, moisture, light, heat, and electric current remains a critical obstacle to practical applications, especially for the most commonly studied 3D methylammonium lead trihalide perovskites. In this respect, low-dimensional perovskite-related organic metal halide hybrids with unique optical and electrical properties[9–14] and better ambient air stability[15] may offer an attractive remedy.[16–24] The integration of various organic cations and metal halides offers the opportunity to obtain a large family of organic metal halide hybrids with low dimensional structures at the molecular levels from 2D to 1D and 0D where the dimensionality describes the connectivity of metal halide specie. For 1D organic metal halide hybrids, metal halide chains/wires/tubes are insulated and surrounded by protective organic cation layers,[20,25–28] suggesting that they may be inherently resilient against various environmental factors. Moreover, in these 1D structures, high electrical conductivity is expected along the chains while the inter-chain charge transport should be suppressed.

In addition to the environmental/chemical stability, an issue of particular relevance to the hybrid perovskite-related materials is their stability against sustained photo illumination and electrical biases. The environmental/chemical instability may be addressed via device encapsulation, the use of buffer layers, and the optimization of perovskite compositions, whereas the photo and electrical instabilities tend to be inherent to the materials and/or devices, and its mitigation requires a microscopic understanding of the charge activation and transport. In 3D halide perovskites, the electrical instability often manifests in the activation and migration of mobile ions due to applied bias or photo illumination, which is sometimes irreversible. Ion migration is therefore relevant to broad optoelectronic applications and has been found to be ubiquitous in halide perovskite solar cells and LEDs; it is believed to be a primary mechanism for the current-voltage (*I-V*) hysteresis in halide perovskite-based devices.[29–32] The hysteresis may be present in the dark, but usually becomes much more pronounced in the presence of photo illumination,[33] and it shows obvious rate dependence, diminishing with decreasing rate of bias scanning.[34–37] The ion migration and diffusion in metal halide perovskites has been probed at microscopic levels through tracking the interdiffusion of different halide anion using optical approaches.[38–40] In some cases, the current-time (*I-t*) measurements exhibit exponential decay or capacitive behavior in various bias ranges.[41,42] Thus far, the electrical hysteresis has been examined almost exclusively in two-terminal (2T) *vertical* junction devices based on the 3D perovskites, including solar cells and LEDs. Such devices comprising a thin active perovskite semiconductor layer sandwiched between two metal electrodes present several complications in elucidating the *intrinsic* physical origins of the hysteresis. Firstly, in common operating conditions, devices were placed under high electric field biases, which often result in irreversible chemical changes in the devices. Furthermore, the metal/perovskite contacts are expected to play prominent, even dominant, roles in the charge transport in the 2T devices, and ion migration can significantly modify the contacts. In order to elucidate the intrinsic ion dynamics of the halide perovskites, including the microscopic mechanism and impacts on the material properties, *4T lateral transport* should be measured in order to eliminate the complications and ensure access to all bias ranges. The chemical instability in 3D halide perovskites makes it extremely challenging to systematically conduct such measurements and obtain reproducible results. In this regard, 1D organic metal halide hybrids offer intriguing promise due to their much-enhanced inherent chemical stability. So far, however, reports of the charge transport measurements on 1D metal halide hybrids have been scarce. Among them, photoconductivity of polycrystalline pellets and arrays of single-crystalline needles of 1D Pb-halides and their derivatives were demonstrated.[43,44] More recently, large-bias space charge limited current (SCLC) was measured on individual single-crystalline needles of 1D Pb-bromide,[45] which yields large room-temperature carrier mobility for the material.

In this work, we focus on the intrinsic *I-V* characteristics and *V*(*t*) behavior in single-crystalline 1D organic metal halide hybrids, elucidated from 4T measurements with constant current biases. 4T *I-V* measurements were performed on individual single-crystalline needles of 1D (R)-α-methylbenzylammonium lead triiodide ((R-α-MBA)PbI$_3$). The resulting *I-V* curves exhibit a C-shaped feature characterized by an initial 'negative' differential resistance region followed by an approach to a linear region. The C-shape depends on the rate and direction of the *I-V* sweep, whereas the slope of the linear section stays constant. The two features hence collectively produce a well-defined, rate-dependent *I-V* hysteresis in loop measurements. Remarkably, the unique *I-V* hysteresis can be fully reproduced with numerical calculations based on a single exponential dynamic with a time constant of 2.0 s, which is verified via 4T *V*(*t*) measurements with constant bias current. The experiments, therefore, constitute direct real-time measurements of the intrinsic ion dynamics in 1D metal halides. The results evidence the essential role of ion migration in the electronic charge carrier transport, which should be accounted for in the design and implementation of optoelectronic devices based on the low-dimensional hybrid metal halides.

**Material Development.** 1D (R-α-MBA)PbI$_3$ single crystals were grown by the inverse temperature crystallization method[46] in which equimolar ratios of PbI$_2$ and R-α-MBA were introduced in HI at 90 °C solution until complete dissolution of PbI$_2$. After cooling down the solution to 40 °C, yellow needle-like crystals were obtained (Figure 1a). Single crystal X-ray diffraction (SCXRD) was used to characterize its structure, which reveals that (R-α-MBA)PbI$_3$ crystallizes in the chiral space group of P 2$_1$ 2$_1$ 2$_1$ within the orthorhombic crystal system. The 1D structure at the molecular level is depicted in Figure 1b, where inorganic chains containing face sharing lead iodide octahedra ((PbI$_6$)$^{4-}$) (Figure 1c) are isolated and surrounded by chiral organic cations (R-α-MBA$^+$) (Figure 1d). The powder X-ray diffraction (PXRD) pattern of the ball-milled sample is identical to the simulated pattern from the single crystal structure (Figure 1e), suggesting the high purity of developed single crystals, which is further confirmed by the Rietveld refinement results (Figure S1).[47,48]

**Charge transport measurements.** To facilitate 4T charge transport measurements, electrical contacts were made on individual needle-like (R-α-MBA)PbI$_3$ crystals by shadow-deposited Cr/Au electrodes or direct silver paint contacts, which yielded similar results. Prior to detailed 4T *I-V* and *V(t)* measurements, a comparative measurement of 4T and 2T *I-V*'s was performed for each sample under ambient air and light to ensure good sample quality and sufficiently low contact resistances. A representative set of 2T and 4T *I-V*'s evidenced ohmic electrical contacts with relatively small contact resistances (Figure S2). The *I-V* curves were obtained stepwise by measuring the voltage after each instrument-controlled constant $I_{bias}$ was applied for a set amount of time, thus a sweep rate corresponds to a specific wait time (Δ*t*). Figure 2a shows a set of *I-V*'s taken with Δ*t* of 0.15 s, 0.25 s, 0.50 s, 1.00 s, 2.00 s and 10.0 s, with $I_{bias}$ starting at -10 nA in steps of 1 nA until +10 nA. The measurements shown were taken under ambient conditions, and the effect of laser illumination will be presented later. For fast sweep rates ($\Delta t$ < 10 s), the *I-V*'s show the C-shaped behavior before transitioning to a linear region. Notably, it took ~ 6 s to reach the linear regime *regardless of the sweep rate*. For $\Delta t$ of 0.15 s and 0.25 s, there is no distinct linear region within the range of the applied biases because the total time to take the entire *I-V* was 3.15 s and 5.25 s respectively. In comparison, for $\Delta t$ of 10 s, the *I-V* is essentially linear. Figure 2b shows a comparative set of *I-V*'s in the opposite sweep direction with a starting current of +10 nA. Clearly, the *I-V*'s are antisymmetric with respect to opposite current sweep directions for all sweep rates. The observations indicate that there is *a single reversible dynamic with a well-defined time constant* governing the charge transport in this material.

*I-V* loop measurements lend further support to the conclusion. A set of *I-V* loops for $\Delta t$ of 0.5 s, 1.0 s and 10 s are shown in Figure 2c. The *I-V* loops were measured with a sequence of $I_{bias}$ applied from -10 nA → 10 nA → -10 nA → 10 nA in steps of 1 nA, the 'start' in the figure indicates the starting point of a loop measurement. A rate-dependent hysteresis is evident. Although current-voltage hysteresis has been widely observed in 3D perovskite solar cells,[34–37,49] it is important to note that their specific characteristics, including the magnitude, shape, and rate dependence, vary greatly among different devices even for the same materials. In contrast, the *I-V* hysteresis here reflect exclusively the intrinsic charge dynamics in the 1D metal-halide, and *all* details of the hysteretic *I-V*'s can be reproduced based on the retardation caused by a single exponential dynamic, as discussed in detail below. Consequently, the *I-V* hysteresis is diminished with decreasing sweep rate; at a long $\Delta t$ (e.g., 10 s), the retardation is effectively minimized, resulting in an approximate linear *I-V* of negligible hysteresis. We emphasize that for the cases of 0.5 s and 1.0 s wait times, the system never reaches a steady state in the entire *I-V* measurement, i.e., there is always a finite ionic current even in the linear sections of the *I-V* curve, and the magnitude of the ionic current depends on the history.

In order to directly verify the exponential dynamic and determine the time constant, we performed 4T *V*(*t*) measurements under several different current biases, in which a constant bias current was applied while the voltage was being continuously monitored. A set of results are shown in Figure 3a. The time 0 is chosen as when $I_{bias}$ is applied. Each *V*(*t*) exhibits an exponential increase from *t* = 0 and approach to a constant value. A common underlying dynamic is evidenced by the fact that all *V*(*t*) can be collapsed (Figure 3b) and fitted to an exponential function,

$$V(t) = V_0(1 - e^{-\frac{t}{\tau}}), \tag{1}$$

and the fittings, represented by the solid lines, yield a single time constant of $\tau$ = 2.049 (±0.086) s. The results are direct evidence that within the experimental bias range, the charge dynamic is independent of the applied current, and after reaching the steady state, the charge transport in the crystal is essentially ohmic. On the other hand, the resulting $V_0$, the steady-state voltage, shows a slight decrease with $I_{bias}$, which is apparent in Figure 3b. We note that, mathematically, Equation 1 is the solution of the voltage across a capacitor and a resistor (RC) in parallel under a constant current bias. Based on the overall experimental observations and the material system involved, we conclude that the physical origin behind Equation 1 is ionic, rather than electronic, in nature. In particular, the long response times (minutes) of the electrical transport to laser illumination (shown in detail below) are conclusive evidence for an ionic origin.

The experimental *V*(*t*) dynamics provided a basis for precise numerical simulation of the complex rate-dependent *I-V* curves in Figure 2: Each voltage in an *I-V* curve can be calculated from Equation 1, based on the specific Δ*t* and expected $V_0$ from the applied current and sample resistance ($R \cdot I_{bias}$). Details of the numeric simulation process are described in Figure S3. The solid lines in Figure 3c are the numerically simulated *I-V* curves based on the exponential dynamics and conditions for measurements shown in Figure 2b; the simulations show good agreement with the corresponding experimental data.

We now turn to the underlying physical process for the distinct charge dynamics. We posit that reversible field-induced ion migration and accumulation is responsible for the charge transport behavior. In this material, the likely mobile ions are the iodide ions.[50–53] A schematic depiction of the ion migration and accumulation processes for the situation of a constant current bias is shown in Figure 4a: Prior to the application of the bias current, the mobile ions were evenly distributed across the sample. As soon as the bias current is applied, the ions begin to migrate, resulting in an ionic current, $I_{ion}$. The total current inside the crystal, maintained to be constant, is the sum of the ionic and electronic current: $I_{bias} = I_{ion} + I_e$. The current is predominantly ionic at the beginning and decreases exponentially with time as the ions accumulate at the anode. The ionic current can then be represented by:

$$I_{ion}(t) = I_{bias} e^{-\frac{t}{\tau}}. \qquad (2)$$

For $t \gg \tau$, the ionic current approaches 0 as a steady state is reached, and the current inside the crystal becomes purely electronic, as illustrated in the bottom panel of Figure 4a. Moreover, because the ionic conductance is negligible in comparison to the electronic conductance, $G_{ion} \ll G_e$, the sample resistance is essentially the electronic resistance, $R \approx R_e$, and the electronic current can be determined as, $I_e(t) = V(t)/R$, where $V(t)$ is the measured voltage at time $t$. Hence, in our setup of constant current measurement, the ionic current can be determined as

$$I_{ion}(t) = I_{bias} - I_e(t) = I_{bias} - V(t)/R, \qquad (3)$$

where $R$ can be evaluated from the slope of the linear sections of the *I-V* curves, which is not exactly the same as the steady state value of $V(t)/I_{bias}$, we speculate the discrepancy came from the different bias conditions in *I-V* and $V(t)$ measurement. Figure 4b shows the calculated $I_{ion}$ using Equation 3 based on the data in Figure 3a. Each $I_{ion}$ follows a similar exponential decrease toward zero regardless of the value of $I_{bias}$; moreover, each $I_{ion}$ (*0*) correlates with their respective $I_{bias}$. Alternatively, $I_{ion}$ (*t*) can be normalized by their respective $I_{bias}$, which collapses all the data into a single exponential curve of $\tau = 2.0$ s, as shown in the inset.

The reversible ion migration and accumulation process also provides a natural physical mechanism for interpreting the observed peculiar rate-dependent *I-V* characteristics, which we illustrate in Figure 4c. In the absence of any ion migration, a linear electronic *I-V* is expected, which passes through the origin and whose slope can be derived from the linear section of the measured *I-V* (or equivalently, from the *I-V* of large $\Delta t$), as represented by the dashed line. For each applied current $I_{bias}$, we can then calculate the voltage expected in the absence of an ionic current ($V_{cal}$) and the difference with the experimental value ($V_{exp}$), $\Delta V = V_{cal} - V_{exp}$. Based on the picture described earlier, $\Delta V$ is caused by the presence of the ionic current, whose magnitude is $I_{ion} = \Delta V/R$, where $R$ is essentially the electronic resistance and can be taken from the slope of the linear region of the *I-V*. Figure 4d shows the calculated $\Delta V/R$ from the experimental *I-V*'s in Figure 2a and 2b, open and solid symbols represent the data calculated from *I-V*'s of different $I_{bias}$ sweep directions. Several features are worth noting: i) The $I_{ion}$ (*t*) do not follow a single common exponential dependence because of the stepwise change of $I_{bias}$ in the *I-V* measurement. ii) For each *I-V*, there is a well-defined point at which $I_{ion}$ changes sign due to the change of $I_{bias}$. Such a crossover point is indicated in Figure 4c. iii) The experimental I-V curves show a voltage offset at $I_{bias} = 0$, which decreases with decreasing sweep rate (increasing $\Delta t$). iv) Finally, $I_{ion}$ saturates at a *nonzero* value which decreases with Δ*t*. This manifests in the *I-V* as a constant finite shift, Δ*V*, in the linear region, as indicated in Figure 4c, and as the Δ*t*-dependent nonzero saturation values of $I_{ion}$ (*t*) indicated in Figure 4d.

In order to elucidate the effect of photoexcitation on the ion dynamics, we performed *I-V* and *V*(*t*) measurements under laser illumination. Figure 5a shows the *I-V* for Δ*t* = 0.5 s under green laser (532 nm) illumination for a (R-α-MBA)PbI$_3$ crystal. An *I-V* curve was first taken in the ambient condition, and the same *I-V* measurement was repeated immediately (0) and 1, 2, 3 minutes after the laser was turned on. Δ*t* is set at 0.5 s, so each *I-V* sweep took about 11 s, which left 49 s of recovery time between any *I-V* measurements, wherein the laser was left on, but no bias current was applied to the sample. The laser was then turned off, and an *I-V* sweep was taken immediately. Finally, 10 minutes after the laser was turned off (under ambient light), the last *I-V* curve was taken, allowing ample sample relaxation time to ascertain the reversibility of photo-activated and field-assisted ion migration. A similar set of *I-V* measurements was performed under blue laser (405 nm) illumination with the same crystal, yielding qualitatively similar results (Figure S4). Most notably, upon laser illumination, the *I-V* curve shows a more pronounced initial C-shape, which further increases and then appear to saturate with increasing illumination time. The observation indicates larger magnitude of the ionic current and longer time constant, which is consistent with enhanced density but decreased mobility for the mobile ions with laser illumination. Moreover, the slope of the linear section of the *I-V* curve clearly decreases with laser illumination, suggesting an increase of the (electronic) sample resistance, probably due to increased ionized impurity scattering.

The *I-V* results and their interpretation are well-corroborated by *V*(*t*) measurements under laser illumination. Figure 5b shows the results with constant current bias of $I_{bias}$ = 10 nA. The data are fitted to Equation 1, as represented by the solid lines. The resulting time constants for ambient, and blue and green laser illumination are 1.97 s, 2.79 s, and 3.91 s, respectively; concomitantly, the saturation voltage, $V_0$, increases from 5.20 V to 7.31 V and 10.35 V. At this point, it is not clear whether the differences between the green and blue laser illumination originate from the different wavelengths or intensities; targeted experiments are needed to clarify this issue. The inset shows the normalized $I_{ion}$ calculated from the $V(t)$ data, which exhibits an obvious increase of the time constant with photoexcitation. Using the method depicted in Figure 4c, $I_{ion}$ can also be calculated from the $I$-$V$ data in Figure 5a; the results are shown in Figure 5c. Both the enhancement of the magnitude of $I_{ion}$ and the increase in the time constant by the laser illumination are evident. Moreover, we note that the initial $I$-$V$ taken under ambient condition and that taken 10 minutes the laser was turned off are virtually identical, suggesting that the photo-activation of mobile ions and field-assisted ion migration are full reversible in this material.

1D orgainc metal halide (R-α-MBA)PbI$_3$ single crystals exhibit excellent chemical stability, which enables highly reproducible time-dependent electrical measurements. We demonstrate that four-terminal $I$-$V$ and $V(t)$ measurements are capable of yielding real-time ionic current, and the experiments identify a single exponential ion dynamic that underlies the charge transport in the material. We further examine the effects of photoexcitation on the ion dynamics and charge transport, the results indicate that the photoexcitation increases the mobile ion density while decreasing the mobility. Both the photoexcitation of mobile ions and field-assisted ion migration are found to be fully reversible. The intrinsic ion dynamics in 1D (R-α-MBA)PbI$_3$ elucidated from this work may provide valuable new insights into the dynamics of ion migration in halide perovskites in general. Much like the understanding of the intrinsic physical properties of semiconductors held the key to the realization of functional semiconducting devices, the definitive elucidation of the intrinsic ion dynamics in the halide perovskites is instrumental to the modeling and detailed understanding of related device properties, particularly the *I-V* hysteresis in solar cells and LEDs, and provide design guidelines for their mitigation and optimization.

## Supporting Information

Experimental details of material preparation, crystal growth and characterization; powder XRD patterns; 2T and 4T measurements comparison; details of numerical simulation and comparison between experimental results; *I-V* and $I_{ion}(t)$ under blue laser illumination; etc.

## Acknowledgment

We thank Moien Adnani for his assistance with the references, and Michael Worku, Zihan Zhang, and Chenkun Zhou for technical assistance and discussions. We acknowledge financial support from the National Science Foundation via grants DMR-1905843 (P.X.), DMR-2204466 (B.M.), and DMR-2110814 (H.G.).

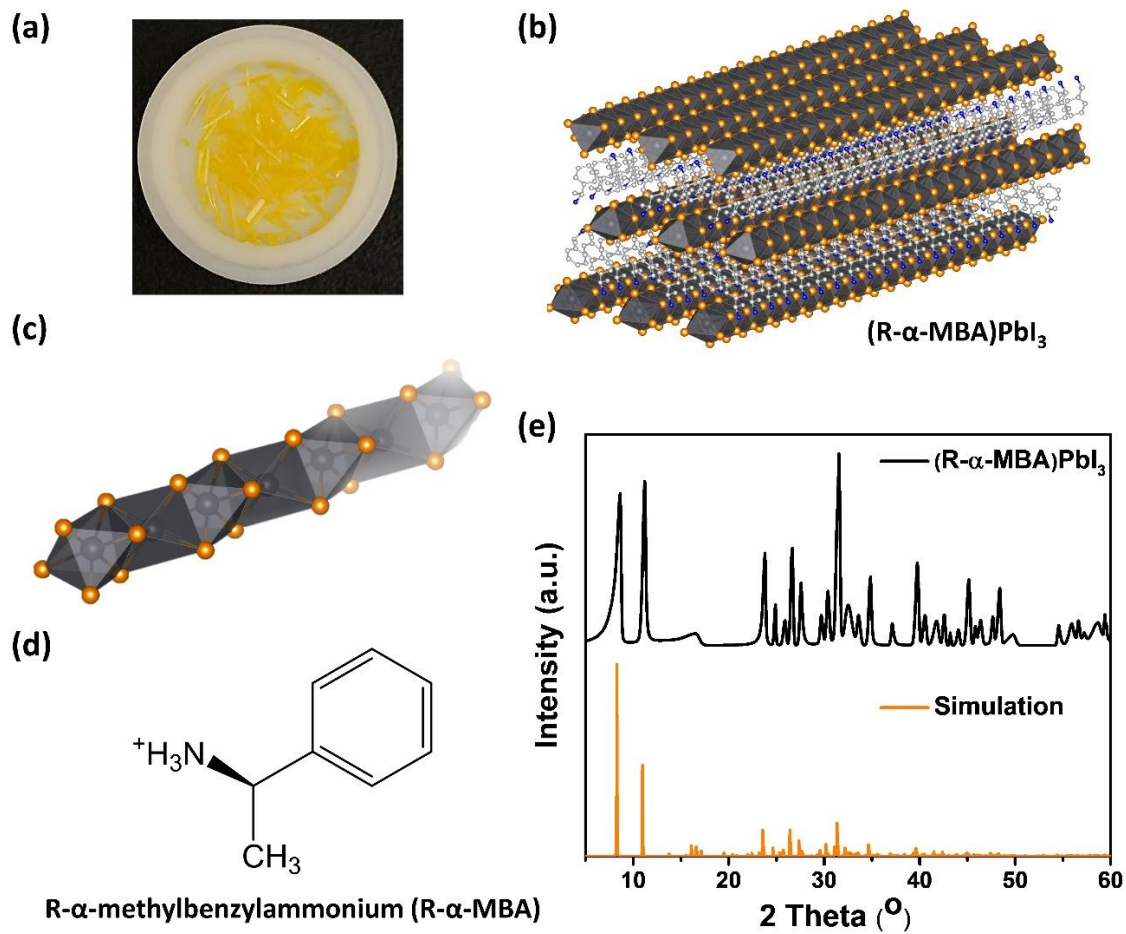

**Figure 1** (a) An optical image of 1D (R-α-MBA)PbI$_3$ single crystals under ambient light. (b) Crystal structure of (R-α-MBA)PbI$_3$ with black, orange, light grey, and blue spheres denoting Pb, I, C, and N atoms, respectively, (H atoms are omitted for clarity). (c) Side view of an individual lead iodide chain containing face sharing octahedra. (d) Molecular structure of R-α-MBA cation. (e) Powder XRD patterns of (R-α-MBA)PbI$_3$ and a simulated pattern.

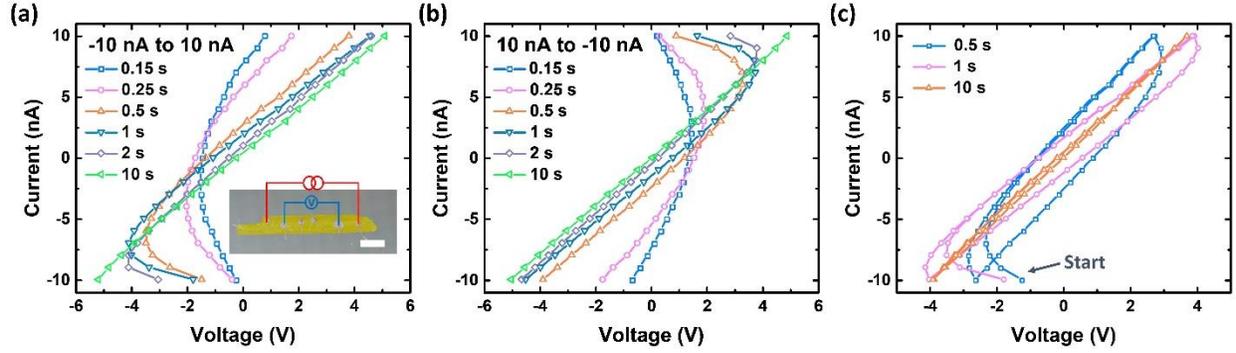

**Figure 2** Four-terminal *I-V* curves in the 1D metal-halide (R-α-MBA)PbI$_3$. (a) *I-V*'s taken with different wait times upon application of bias current from -10 nA to 10 nA in steps of 1 nA. Inset: optical image of a wired crystal and a schematic of the measurement setup. The scale bar is 1 mm. (b) A comparative set of *I-V*'s taken with opposite current sweep direction from +10 nA to -10 nA. (c) Full loop *I-V* measurements for three different current sweep rates, the 'start' indicates the starting point of each loop measurement.

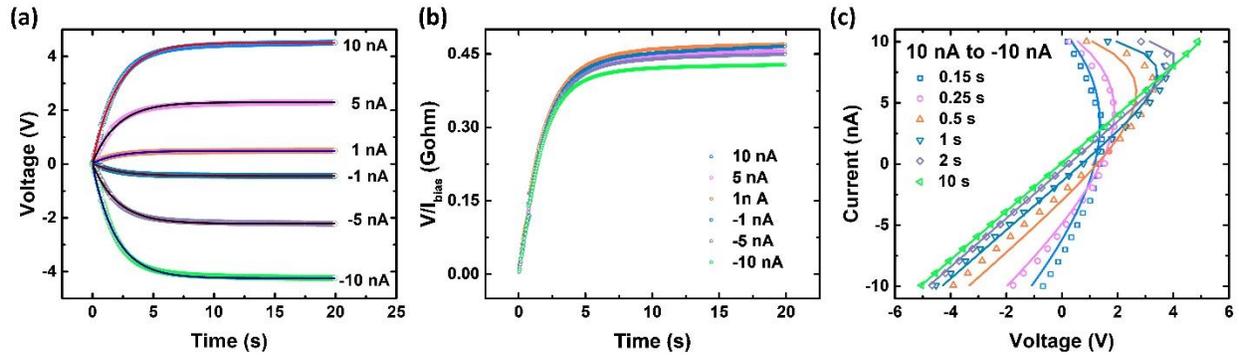

**Figure 3** Single exponential dynamic from *V(t)* measurements. (a) Four-terminal *V(t)* under various constant $I_{bias}$. The solid lines are fittings to Equation 1, which yield a time constant $\tau =$ 2.049 (±0.086) s. (b) *V(t)* normalized by their respective applied bias current. (c) Numeric simulation of the *I-V*'s (solid lines) from Equation 1, based on the wait time and expected $V_0$ from the applied current and sample resistance.

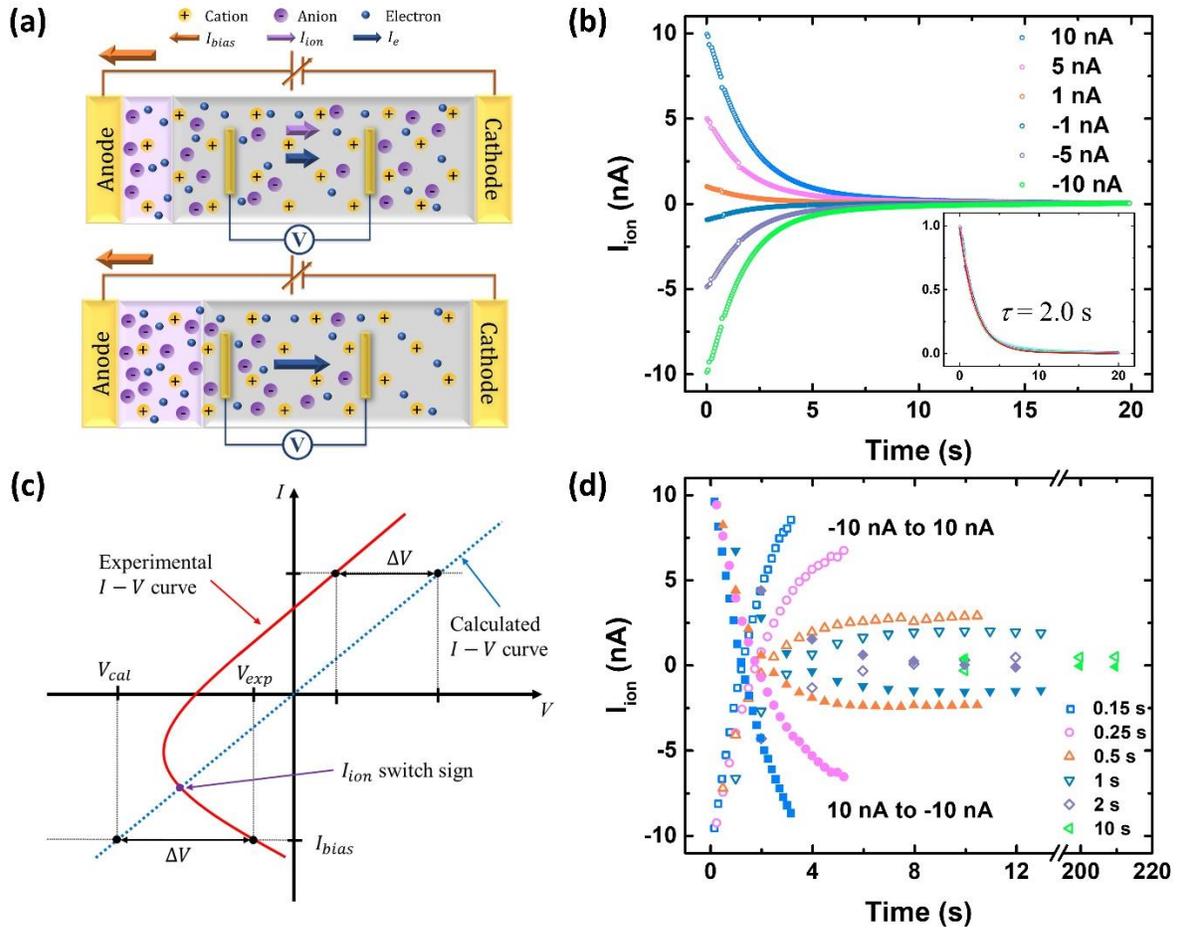

**Figure 4** Real-time ionic current from $V(t)$ and $I$-$V$ measurements. (a) Schematics of the ion migration and accumulation processes. The top panel depicts the initial state upon application of a bias current ($I_{bias} = I_{ion} + I_e$), and the bottom panel represents the steady state with ion accumulation ($I_{bias} = I_e$). (b) $I_{ion}(t)$ calculated with Equation 3 based on the $V(t)$ data in Figure 3a. The resulting $I_{ion}(t)$ follow the same exponential decrease to 0 regardless of the value of $I_{bias}$. Inset: $I_{ion}(t)$ normalized by their respective $I_{bias}$. (c) Schematic illustration of the extraction of rate ($\Delta t$)-dependent $I_{ion}(t)$ in an $I$-$V$ measurement. $I_{ion}$ is calculated from the voltage shift from the linear electronic $I$-$V$ (dashed line): $\Delta V = V_{cal} - V_{exp}$. (d) Extracted $I_{ion}(t)$ from the $I$-$V$'s in Figure 2a and 2b, open and solid symbols represent the data calculated from $I$-$V$'s of -10 nA to 10 nA and 10 nA to -10 nA sweeps respectively.

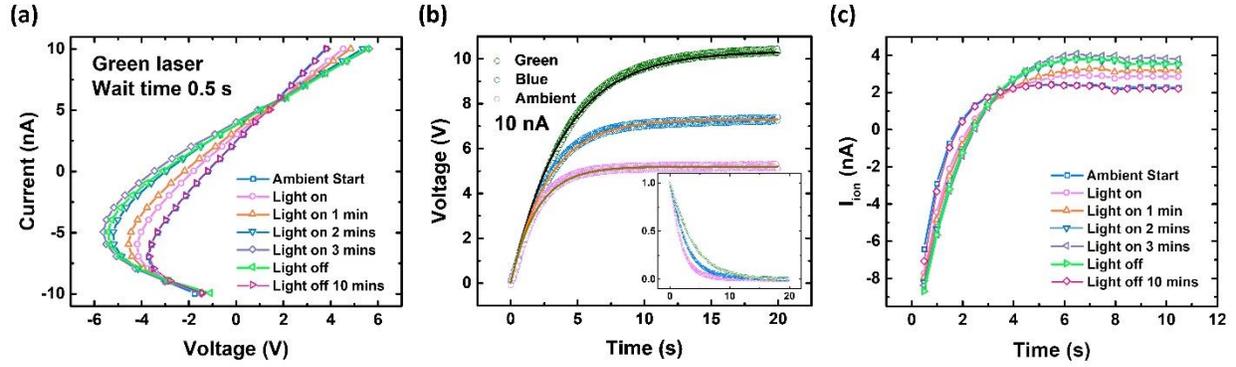

**Figure 5** Effects of laser illumination. (a) *I-V* curves for a 1D (R-α-MBA)PbI$_3$ crystal under ambient light and green laser (532 nm) illumination. The results reveal that the laser illumination simultaneously increases the magnitude of the ionic current and the time constant of the ion dynamics. (b) Four-terminal *V*(*t*) for the same crystal under ambient light and blue and green laser illumination. The solid lines are fittings to Equation 1, which result in $\tau$ = 1.97s, 2.79 s, and 3.91 s for ambient condition and green and blue laser illumination, respectively. Inset: Normalized $I_{ion}$(*t*) from the *V*(*t*) data. (c) $I_{ion}$(*t*) calculated from the *I-V* in Figure 5a using the same method described in Figure 4c.

Supporting Information for:

# Intrinsic ion migration dynamics in a one-dimensional organic metal halide hybrid

Zhenqi Hua[1], Azza Ben-Akacha[2], Qingquan He[2], Tianhan Liu[1], Gillian Boyce[1], Margaret van Deventer[1], Xinsong Lin[2], Hanwei Gao[1], Biwu Ma[2, *], and Peng Xiong[1, *]

*1Department of Physics, Florida State University, Tallahassee, Florida 32306, USA*

*2 Department of Chemistry and Biochemistry, Florida State University, Tallahassee, Florida 32306, USA*

*Email: bma@fsu.edu, pxiong@fsu.edu*

# 1. Material Synthesis and Characterization

**Materials.**

Lead iodide ($PbI_2$, 99.999%), (R)-α-methylbenzylamine (R-α-MBA, 98%), hydroiodic acid (HI, 57%), hypophosphorous acid solution ($H_3PO_2$, 50 wt. % in $H_2O$) were purchased from Sigma Aldrich and were used without further purification.

**1D (R-α-MBA)$PbI_3$ crystal growth.**

1D chiral crystals were prepared by adding 2 mmol of lead iodide and (R)-α-methylbenzylamine into 6 mL of hydroiodic acid and 0.5 mL of hypophosphorous acid in a round bottom flask. The mixture was stirred at 90 ºC until $PbI_2$ completely dissolved then left to cool down slowly to 40 ºC. Long yellow crystals were obtained in few hours. The crystals were filtrated and dried under reduced pressure for further use.

**Powder XRD.**

The powder X-ray diffraction (XRD) analysis was performed on Rigaku MiniFlex X-ray diffractometer, equipped with a D/tex Ultrax detector and a copper Kα radiation source ($\lambda$ = 0.154 nm, 40 kV, 15 mA). The diffraction pattern was scanned over the angular range of 5−80° (2 Theta) with a step size of 0.05° at room temperature. The analysis of PXRD pattern was performed with SmartLab Studio II software. Simulated powder pattern was calculated by Mercury software using the crystallographic information file from this reference[1].

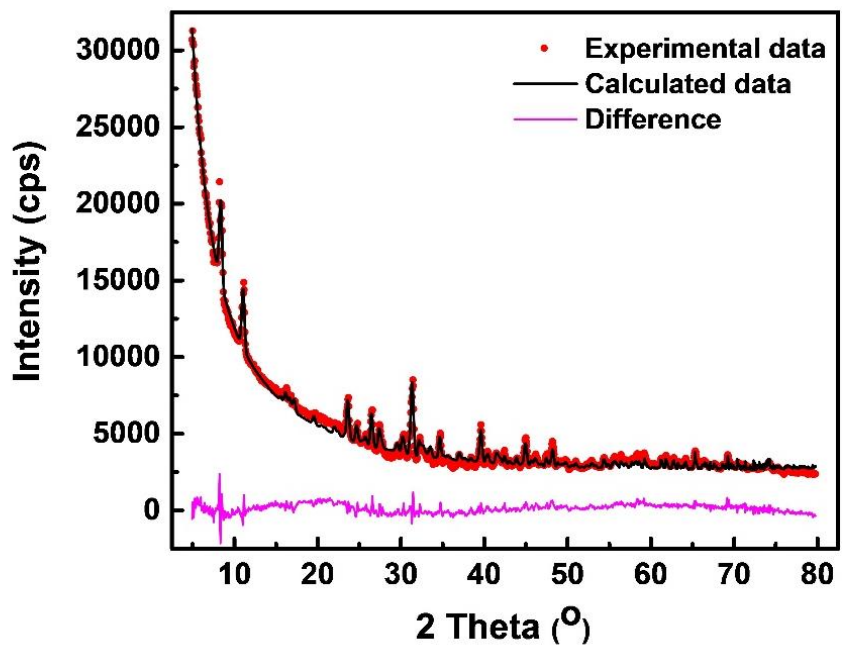

**Figure S1** Plot of experimental powder XRD patterns and the Rietveld refinement data for the synthesized (R-α-MBA)PbI$_3$. The peaks from the experimental data collected from ground crystals are consistent with the ones extracted from the reported cif file[1].

## 2. Comparison of 2T and 4T Measurements

To ensure adequate sample quality and ohmic low-resistance contacts, a comparative 2T and 4T I-V measurement was performed prior to each full set of 4T I-V and V(t) measurements. Figure S2 shows the 2 T and 4T I-V curves for one of the two samples presented in the main text. $I_{bias}$ was ramped from -10 nA to 10 nA in steps of 1 nA, with a wait time Δ$t$ of 1 s. The 2T and 4T I-V curves show the same behavior, with quantitatively similar C-shapes and a slight difference in the slopes of the linear sections, corresponding to 2T resistance of 623 MΩ and 4T resistance of 511 MΩ. Therefore, the contacts are ohmic in the bias range, and have a resistance of about 10% of the sample resistance.

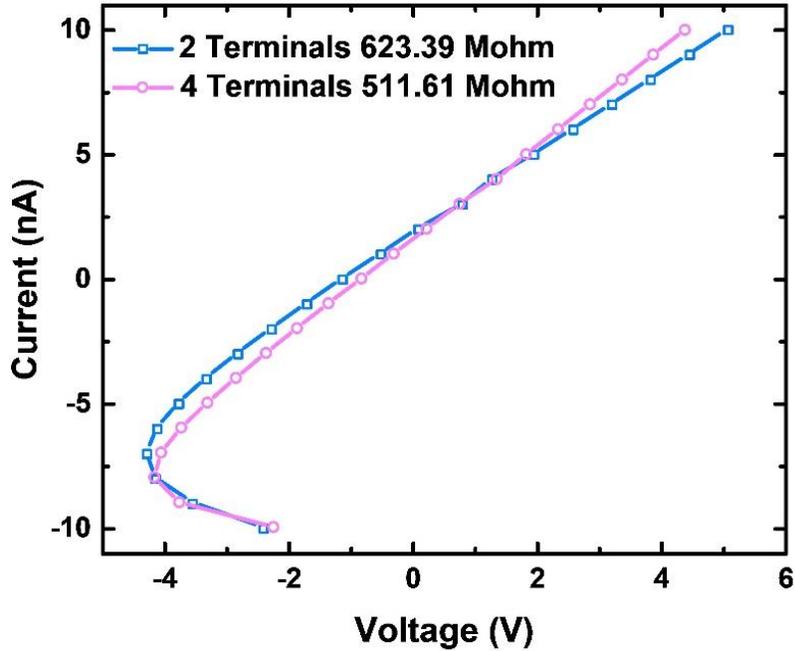

**Figure S2** Comparative 2T and 4T I-V measurements for a 1D (R-α-MBA)PbI$_3$ crystal, evidencing low-resistance ohmic contacts on the crystal.

## 3. Numeric Simulation of I-V and V(t)

The numerical calculation was performed via MATLAB. It stimulates the results of the specific I-V measurement and calculates the V(t). The calculated I-V curves in Fig 3c (solid lines) are obtained this way. The block diagram of the calculation is shown in Fig S3a; by inputting the resistance $R$, time constant $\tau$, and wait time $T$, a full set of simulations can be implemented. The parameter *Flag* represents the ion accumulation direction, and the *Step* is the current ramp step size, $t_0$ is a parameter describing the amount of ion accumulation at each stage, which also ensures the continuity of the function. Specifically, in the case of Fig. 3c, $\tau$ is fixed at 2 s, and $R$ is chosen as the slope of the I-V curve for $\Delta t = 10$ s.

To fully simulate our I-V measurements via stepwise sweep of $I_{bias}$ and facilitate direct comparison with the experimental results, a V(t) measurement was performed in which $I_{bias}$ was manually changed by 1 nA at every $\Delta t$. The experiment was intended to obtain the V(t) in the process of an actual I-V measurement. The results of $\Delta t$ of 1 s and 10 s are shown in Figs. S3b and S3c respectively. The solid red squares are experimental data from the corresponding I-V curves. For the case of $\Delta t = 1$ s, the calculation and the I-V measurement data show excellent agreement, while the discrepancy by the measured V(t) via manual control of $I_{bias}$ is probably due to a delay in the manual control.

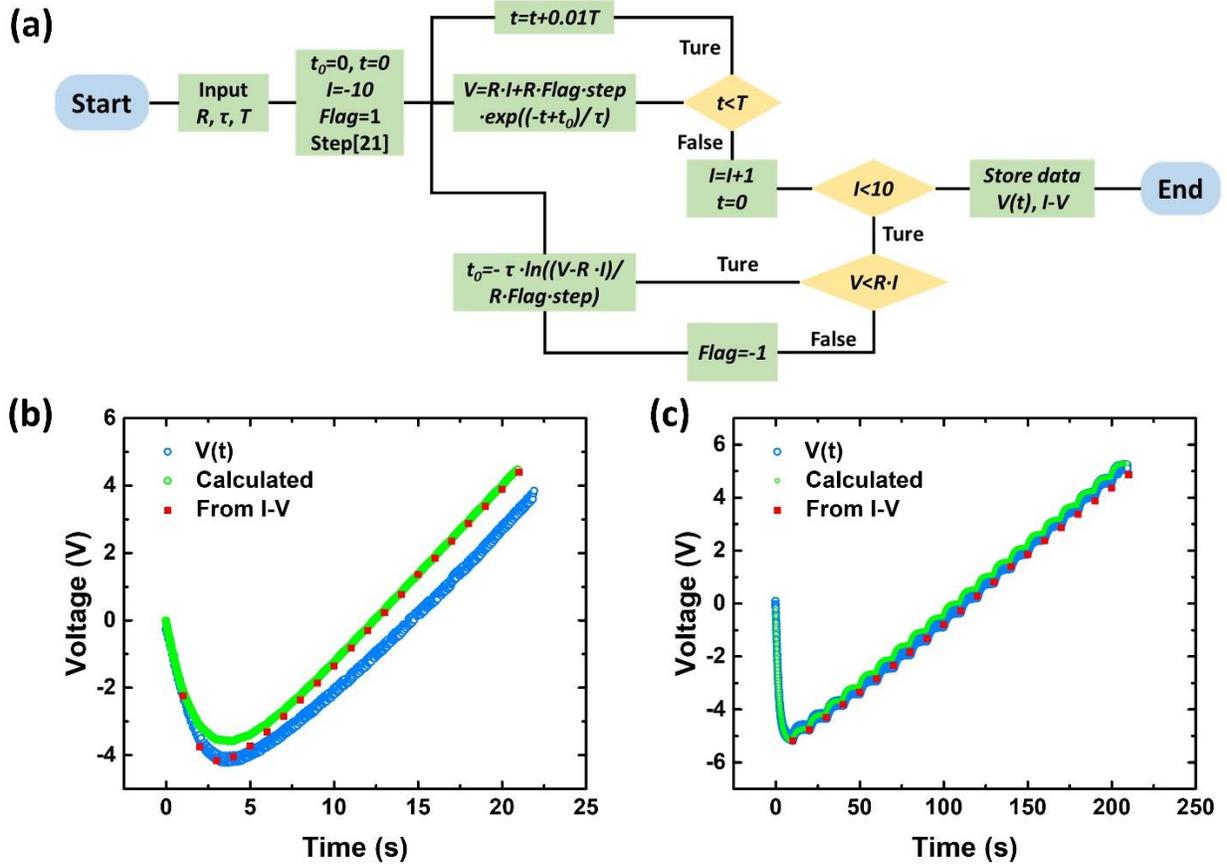

**Figure S3** (a) Block diagram of the process of numeric simulation. (b), (c) V(t) measured under manual change of $I_{bias}$ in steps of 1 nA, in comparison to the calculation results and data from corresponding I-V curves for $\Delta t$ of (b) 1 s, and (c) 10 s.

## 4. I-V under Blue Laser Illumination

The I-V and calculated $I_{ion}$ with time under the blue laser (405 nm) illumination are shown in Figs. S4a and S4b, respectively. The results are qualitatively consistent with those with green laser illumination shown in the main text. Quantitatively, the variation of the C-shape with blue laser illumination is less dramatic and the increase of $I_{ion}$ is less than those of the green laser illumination. Concomitantly, the increase of (electronic) sample resistance and the time constant also less for the blue laser illumination less than those for the green laser illumination.

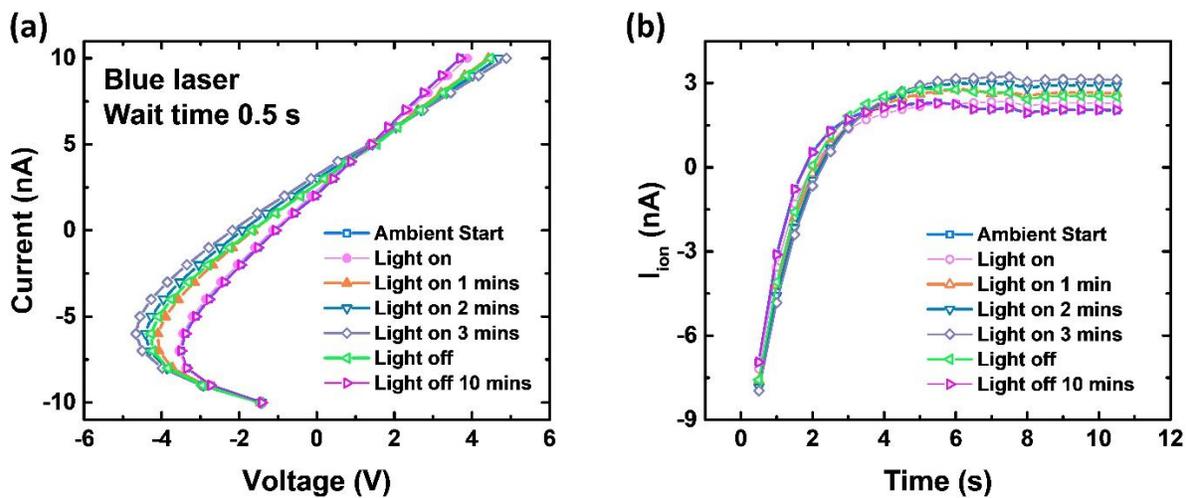

**Figure S4** (a) 4T I-V curves under blue laser (405 nm) illumination. (b) $I_{ion}(t)$ calculated from (a).